\documentclass[final,3p,times,twocolumn]{elsarticle}
 \biboptions{comma,sort&compress}
\usepackage{here}
 \usepackage{graphicx}
  \usepackage{epsfig}
\def\nin{\noindent}
\def\beq{\begin{equation}}
\def\eeq{\end{equation}}
\def\bea{\begin{eqnarray}}
\def\eea{\end{eqnarray}}

\journal{Nuc. Phys. (Proc. Suppl.)}

\begin{document}

\begin{frontmatter}



\title{\boldmath Puzzles in charmonium decays}

 \author[label1]{Qiang Zhao\corref{cor1}}
  \address[label1]{Institute of High Energy Physics, Chinese Academy of Sciences, Beijing 100049, China,
\\
Theoretical Physics Center for Science Facilities, Chinese Academy
of Sciences, Beijing 100049, China} \cortext[cor1]{Speaker}
\ead{zhaoq@ihep.ac.cn}

% \author[label1,label2]{Marina Nielsen,\corref{label3}}
%  \address[label2]{Instituto de F\'{\i}sica, Universidade de S\~{a}o Paulo,
%C.P. 66318, 05389-970 S\~{a}o Paulo, SP, Brazil}
%\cortext[label3]{Supported by FAPESP within the France-Brazil program.}
%\ead{mnielsen@if.usp.br}
%\author{}

%\address{}

\begin{abstract}
\noindent

The open charm effects via intermediate hadron loop transitions seem
to play a crucial role in the understanding of several existing
``puzzles" in charmonium exclusive decays, such as the $\psi(3770)$
non-$D\bar{D}$ decays, and ``$\rho\pi$ puzzle" etc. In the
charmonium energy region, non-perturbative mechanisms could be still
sizeable, and as a consequence the intermediate hadron loop
transitions also provide a mechanism for the helicity-selection-rule
(HSR) violation. We report our recent progress on those existing
puzzles.

\end{abstract}


\end{frontmatter}

% \linenumbers

\section{Helicity selection rule violations and open charm effects in charmonium decays}
%\label{}
\nin
%%%%%%%%%%%%

The exclusive decays of heavy quarkonium have been an important
platform for studying the nature of strong interactions in the
literature~\cite{Brodsky:1981kj,Chernyak:1981zz,Chernyak:1983ej,Brambilla:2004wf,Voloshin:2007dx,bes-iii}
since the discovery of quantum chromodynamics (QCD). An interesting
issue in charmonium decays is the effects arising from open charmed
meson channels. For charmonia which are close to an open charmed
meson channel, and if they have also strong couplings to the open
channel, one would expect that such an open channel can affect
properties of the charmonia in both spectra and decays. An immediate
example is the $D\bar{D}$ threshold ($m_{D\bar{D}}\simeq 3.72$ GeV),
below and above which the $\psi(3686)$ and $\psi(3770)$ sit closely.
An interesting and nontrivial question here is whether the
$\psi(3770)$ decay is totally saturated by $D\bar{D}$, or whether
there exist significant non-$D\bar{D}$ decay
channels~\cite{Kuang:1989ub,Ding:1991vu,Rosner:2001nm,Rosner:2004wy,Eichten:2007qx,Voloshin:2005sd,Achasov:2005qb}.

Unfortunately, a definite answer from either experiment or theory is
unavailable. In experiment, the $D\bar{D}$ cross section measurement
by CLEO Collaboration suggests that  the maximum non-$D\bar{D}$
branching ratio is about 6.8\%
\cite{He:2005bs,:2007zt,Besson:2005hm}, while BES Collaboration find
much larger non-$D\bar{D}$ branching ratios of $\sim 15\%$ in the
direct measurement of non-$D\bar{D}$ inclusive cross
section~\cite{bes-3770}. In theory, next-to-leading-order (NLO) pQCD
calculation of the $c\bar{c}$ annihilation width for $\psi(3770)$
leads to a maximum of about $5\%$ for the $\psi(3770)$
non-$D\bar{D}$ decay branching ratio~\cite{He:2008xb}, which appears
to favor a relatively smaller non-$D\bar{D}$ branching ratio.
However, as shown in Ref.~\cite{He:2008xb}, the NLO corrections are
the same order as the leading order results. It would jeopardize the
validity of the perturbation expansion on the one hand, and on the
other hand, raises the question about the contributions from
non-pQCD mechanisms.

A relevant issue in exclusive processes is the so-called helicity
selection rule (HSR) \cite{Brodsky:1981kj,Chernyak:1981zz} which
provides a guidance for expectations of perturbative QCD (pQCD)
asymptotic
behaviors~\cite{Brodsky:1981kj,Chernyak:1981zz,Chernyak:1983ej,Brambilla:2004wf,Voloshin:2007dx,bes-iii}
that can be examined in experiment. However, in comparison with the
accumulated data, more and more observations suggest significant
discrepancies between the data and the selection-rule expectations.
As an example,  the decays of $J/\psi \to VP$ and $\eta_c \to VV$
would be suppressed by this rule~\cite{Feldmann:2000hs}. In reality,
they are rather important decay channels for $J/\psi$ and $\eta_c$,
respectively~\cite{Amsler:2008zzb}. One possible reason why the
perturbative method fails here could be that although the mass of
the charm quark is heavy, it is, however, not as heavy as pQCD
demands. Therefore, it is not safe to apply the helicity selection
rule to charmonium decays (One may anticipate that the situation
should be improved in bottomonium decays).

In this proceeding, we report our study of the intermediate meson
loop effects in several HSR-violating channels, which serves as a
solution for our understanding of several puzzling problems, i.e.
$\psi(3770)$ non-$D\bar{D}$ decay, ``$\rho\pi$ puzzle", and
$\chi_{cJ}$'s HSR-violating decays.

%%%%%%%%%%%%
\section{Non-$D\bar{D}$ decay of $\psi(3770)$ }
%\label{}
\nin
%%%%%%%%%%%%

The puzzling situation with the $\psi(3770)$ non-$D\bar{D}$ decay
branching ratio could be a handle for putting pieces of information
together in charmonium exclusive decays. The dominant decay of
$\psi(3770)\to D\bar{D}$ implies possible significant branching
ratios for $\psi(3770)\to \mbox{non}-D\bar{D}$ via a long-range
interaction mechanism. We argue that the intermediate $D\bar{D}$ and
$D\bar{D^*}+c.c.$ rescatterings, which annihilate the $c\bar{c}$ at
relatively large distance by the OZI-rule evading processes, may
provide a natural mechanism for quantifying the $\psi(3770)$
non-$D\bar{D}$ decays~\cite{Zhang:2009kr}. Since $\psi(3770)$ is
above the $D\bar{D}$ threshold, this contribution has an absorptive
part, for which the quantitative results can be pursued.

We investigate the exclusive decay of $\psi(3770)\to VP$, where $V$
and $P$ stand for light vector and pseudoscalar meson, respectively.
This type of transitions is supposed to be suppressed by the HSR at
QCD leading twist similar to $J/\psi\to VP$. However, the non-pQCD
transitions via intermediate meson loop contributions could be a
mechanism evading not only the OZI rule but also the HSR.

An apparent advantage in $\psi(3770)\to VP$ is that we can benefit
from the unique Lorentz structure of antisymmetric tensor coupling.
In principle, all possible transition mechanisms will contribute to
the corrections to the $VVP$ coupling form factor. Thus, we can make
the following parametrization for $\psi(3770)\to VP$:
 \begin{eqnarray}\label{factorization}
 {\mathcal M}_{fi}
&\equiv & i(g_{L}+ e^{i\delta} g_S {\cal F}_S({\vec p}_V)) \nonumber\\
& & \times   \varepsilon_{\alpha\beta\mu\nu} P_\psi^\alpha
\epsilon_\psi^\beta P_V^\mu \epsilon_V^{*\nu}/M_{\psi(3770)}
 \end{eqnarray}
where $g_{L}$ denotes the intermediate meson loop amplitude, and
$g_S$ describes the production of two pairs of $q\bar{q}$ in the
final $VP$ via OZI singly disconnected (SOZI) transitions. The
coupling $g_S$ is related to the short-range contributions. Hence,
we assume that the SU(3) flavor symmetry will connect all the light
$VP$ production channels via
 \begin{eqnarray}
&g_S^{\rho^{0}\pi^{0}}:g_S^{K^{*+}K^{-}}: g_S^{\omega\eta}:
g_S^{\omega\eta'}:g_S^{\phi\eta}:g_S^{\phi\eta'}\nonumber\\
=& 1 :1 : \cos\alpha_P:\sin\alpha_P: (-\sin\alpha_P): \cos\alpha_P \
,
\end{eqnarray}
with the other isospin channels implicated. The angle
$\alpha_P\equiv \theta_P + \arctan(\sqrt{2})$ is $\eta$ and
$\eta^\prime$ mixing angle. A conventional form factor, $ {\cal
F}_S^2({\vec P}_V) \equiv \exp ({-{\vec P}_V^2/{8\beta^2}})$ with
$\beta = 0.5\mbox {GeV}$, is applied for the SOZI transition with
${\vec P}_V$ the final three momentum in the $\psi(3770)$ rest
frame~\cite{close-et-al,Li:2007ky}. The introduction of hadronic
degrees of freedom via the meson loops may lead to a relative phase
$\delta$  between $g_{L}$ and $g_S$ terms.

\begin{figure}[hbt]
\centerline{\includegraphics[width=8.cm]{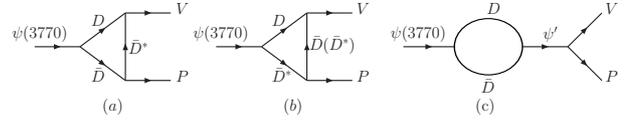}}
%{\epsfig{figure=mpsi2mc.eps,height=70mm}}
\caption{\scriptsize The $t$ [(a) and (b)] and $s$-channel (c) meson
loops in $\psi(3770)\to VP $. \label{fig-1}}
\end{figure}

In Fig.~\ref{fig-1}, the intermediate meson loops recognized as $t$-
and $s$-channel transitions are illustrated. The following effective
Lagrangians are adopted to evaluate the transition amplitudes
related to $g_{L}$,
 \begin{eqnarray}
\mathcal{L}_{\psi D\bar D} &=& g_{ \psi D\bar D}\{D
\partial_\mu {\bar D}-\partial_\mu D\bar{D} \}{\psi^\mu},\nonumber\\
 \mathcal{L}_{\mathcal V D {\bar D}^\ast} &=& -i
g_{\mathcal V D {\bar D}^\ast} \epsilon_{\alpha\beta\mu\nu}
\partial^\alpha \mathcal{V^\beta} \partial^\mu \bar {D^\ast}^\nu D + H.c. ,\nonumber\\
 \mathcal{L}_{\mathcal P D^\ast {\bar D}^\ast} &=& -i
g_{\mathcal P D^\ast {\bar D}^\ast} \epsilon_{\alpha\beta\mu\nu}
\partial^\alpha D^{\ast\beta} \partial^\mu \bar{D}^{\ast\nu}  \mathcal{P} + H.c. ,
 \nonumber \\
 \mathcal{L}_{\mathcal P \bar D D^\ast} &=& g_{D^\ast \mathcal P
 \bar D}\{\bar D
\partial_\mu {\mathcal P}-\partial_\mu {\bar D} \mathcal P\} D^{\ast\mu} +H.c.,
 \end{eqnarray}
where $\epsilon_{\alpha\beta\mu\nu}$ is the Levi-Civita tensor; $
\mathcal P $ and $ \mathcal V^\beta$ are the pseudoscalar and vector
meson fields, respectively.

The charmed meson couplings to light meson are obtained in the
chiral and heavy quark limits~\cite{Cheng:2004ru},
 \begin{eqnarray}
 g_{D^\ast D \pi} = \frac{2}{ f_\pi } g
\sqrt{m_D m_{D^\ast}} , &  &   g_{D^\ast D^\ast \pi} =
\frac{g_{D^\ast D \pi}}{\tilde{M_D}} ,\nonumber\\
g_{D^\ast D \rho} = \sqrt{2} \lambda g_\rho , && g_{D D \rho} =
{g_{D^\ast D \rho}} \tilde{M_D} ,
 \end{eqnarray}
where $f_\pi$ = 132 MeV is the pion decay constant, and
$\tilde{M_D}\equiv \sqrt{m_D m_{D^\ast}}$ sets a mass scale. The
parameters $g_\rho$ respects the relation $g_\rho = {m_\rho /
f_\pi}$~\cite{Casalbuoni:1996pg}. We take $\lambda = 0.56\,
\mbox{GeV}^{-1} $ and  $g = 0.59$ ~\cite{Liu:2006dq,Yan92}.

%%%%%%%%%%%%%%%%%%%%%%%
\nin
%%%%%%%%%%%%%%%%%%%%%%%

 Note that the $t$-channel loops suffer from
divergence~\cite{Zhang:2008ab}. A dipole form factor is thus
introduced to kill the divergence and also compensate the off-shell
effects arising from virtual particle exchanges:
 \begin{equation}\label{ff-loop}
 {\cal F}(q^2)=\left(\frac{\Lambda^2-m_{ex}^2}{\Lambda^2-q^2}\right)^2 \ ,
 \end{equation}
 where $\Lambda \equiv  m_{ex} +
\alpha \Lambda_{QCD}$, with $\Lambda_{QCD} = 0.22$ GeV;  $m_{ex}$ is
the mass of the exchanged meson and $\alpha$ is a parameter to be
determined by experimental data for $\psi(3770)\to J/\psi\eta$. The
detailed formulae, which we skip here, can be found in
Ref.~\cite{Zhang:2009kr}.

We use the measured $\psi(3770)\to J/\psi\eta$ to constrain the form
factor parameter $\alpha$. Note that the momentum carried by the
final state mesons is rather small. This is an indication that the
gluon exchanges are very soft. We hence assume that the
short-distance pQCD processes are strongly suppressed, and the
transition is dominated by the long-distance transition mechanism
via intermediate meson loops. With $BR^{exp}_{J/\psi\eta}=(9.0\pm
4)\times 10^{-4}$~\cite{Amsler:2008zzb}, $\alpha=1.73$ can be
determined and the exclusive $t$-channel contributes $8.44\times
10^{-4}$ to the branching ratio.

By applying the experimental data, $BR_{\phi\eta}=(3.1\pm 0.7)\times
10^{-4}$~\cite{Amsler:2008zzb} and $BR_{\rho\pi}<0.24\%$ with C.L.
of 90\%~\cite{Ablikim:2005cd}, we can constrain the coupling $g_S$
and phase angle $\delta$. Other channels can then be
predicted~\cite{Zhang:2009kr}.

 %%%%%%%%%%%%%%%%%%%%%%%%%%%%%%%
\begin{center}
{\scriptsize
\begin{table}[hbt]
\setlength{\tabcolsep}{1.4pc}
 \caption{\scriptsize    Branching ratios for $\psi(3770)\to VP$
calculated for different mechanisms. The values for $J/\psi\eta$ and
$\phi\eta$ are fixed at the central values of the experimental
data~\protect\cite{Amsler:2008zzb}, and the experimental upper limit
is taken for $\rho\pi$~\protect\cite{Ablikim:2005cd}.}
    {\small
\begin{tabular}{ccc}
\hline
BR$(\times 10^{-4})$  & Total & Exp.  \\
\hline\hline  $J/\psi\eta$           &       $9.0$      &    $9.0\pm 4.0$       \\
\hline  $J/\psi\pi^0$          &  $4.4\times 10^{-2}$   & $<2.8$
\\
\hline  $\rho\pi$              &  $24.0$  &  $<24.0$~\protect\cite{Ablikim:2005cd}\\
\hline  $K^{*+} K^- + c.c$     &  $8.91$  & not seen \\
\hline  $K^{*0} {\bar K}^0+c.c$&  $9.90$  & not seen \\
\hline  $\phi\eta$             &  $3.1$   & $3.1\pm 0.7$ \\
\hline  $\phi\eta^\prime$      &  $3.78$  & not seen \\
\hline  $\omega\eta$           &  $4.69$  & not seen \\
\hline  $\omega\eta^\prime$    &  $0.39$  & not seen  \\
\hline  $\rho\eta$             &  $1.8\times 10^{-2}$  & not seen \\
\hline  $\rho\eta^\prime$      &  $1.0\times 10^{-2}$  & not seen \\
\hline  $\omega\pi^0$          &  $2.5\times 10^{-2}$  & not seen \\
\hline  Sum          &
$63.87$ & - \\
\hline
%\hline
\end{tabular}
} \label{tab-1}
\end{table}
}\end{center} \nin

In Tab.~\ref{tab-1}, the results for all $VP$ channels are listed.
The exclusive branching ratios given by the $t$- and $s$-channels,
and SOZI transitions can be found in Ref.~\cite{Zhang:2009kr}.
Remember that the input channels are $\psi(3770)\to J/\psi\eta$,
$\phi\eta$ and $\rho\pi$, where the experimental upper limit for
$\rho\pi$ suggests a correlation between the SOZI transition
coupling $g_S$ and the phase angle $\delta$. By varying $\delta$,
but keeping the $\phi\eta$ rate unchanged (i.e. $g_S$ will be
changed), we obtain a bound for the sum of branching ratios, $\simeq
(0.41-0.64)\%$. We refer readers to Ref.~\cite{Zhang:2009kr} for the
detailed calculation, and only summarize the conclusive points as
follows:

i) It is interesting to see that the intermediate $D$ meson loop
transitions indeed account for some deficit for the non-$D\bar{D}$
decay. In particular, the $t$-channel transitions illustrated in
Fig.~\ref{fig-1} contribute dominantly to the decay amplitudes. In
contrast, in most cases, the SOZI and $s$-channel transitions are
found rather small~\cite{Zhang:2009kr}.

ii) The sum of those $VP$ channel contributions accounts for the
branching ratio of $(0.41-0.64)\%$ in the $\psi(3770)$
non-$D\bar{D}$ decays. This appears to be a small fraction. However,
notice that $\psi(3770)$ opens to a large number of light meson
decay channels. A sum of all those channels may result in a sizeable
non-$D\bar{D}$ branching ratio. High statistic experiment at
BEPCII~\cite{bes-iii} should be able to measure more exclusive decay
channels of $\psi(3770)$, which will be able to provide a direct
test of our mode calculations.

iii) We stress again that our calculation of $\psi(3770)\to VP$
benefits from the unique property of the $VVP$ coupling. It allows a
reliable constraint on the model parameters. For other exclusive
channels, e.g. $\psi(3770)\to VS$ and $VT$, the vertex couplings
become complicated. Numerical calculations of the intermediate meson
loops would become strongly model-dependent. Qualitatively, the
results for $\psi(3770)\to VP$ provide an idea that how large an
exclusive decay branching ratio would be in the $\psi(3770)$
non-$D\bar{D}$ decays~\cite{Liu:2009dr}.

%%%%%%%%%%%%
\section{Brief comments on the ``$\rho\pi$ puzzle"}
%\label{}
\nin
%%%%%%%%%%%%
As mentioned earlier, the decay of $\psi(3770)\to VP$ also violates
the HSR at pQCD leading twist. Therefore, the intermediate meson
loop transitions provide a non-perturbative mechanism for evading
the HSR. A natural conjecture is that such a mechanism may also play
a role in $\psi^\prime\to VP$. Since the mass of $\psi(3686)$ is
located in the vicinity of the open $D\bar{D}$ threshold, it would
experience much more significant effects from the open channels than
$J/\psi\to VP$. In this sense, the non-perturbative intermediate
meson loop transitions could be correlated with the ``$\rho\pi$
puzzle" in the literature.

As shown in Refs.~\cite{Li:2007ky,Zhao:2006gw}, there exists an
overall suppression on the short-range strong decay strength of
$\psi(3686)\to VP$, not just in $\psi(3686)\to \rho\pi$. Due to this
suppression, the EM transition amplitudes become compatible with the
strong decay amplitudes with which the interferences produce
deviations from the naive expectations based on the SOZI transition
mechanism. To be more specific, due to the interference, the
$\rho\pi$ decay is further suppressed, i.e. causes the so-called
``$\rho\pi$ puzzle". Also, the neutral $K^{*0}\bar{K^0}+c.c.$ has a
much larger branching ratio than the charged one
$K^{*+}K^-+c.c.$~\cite{Amsler:2008zzb}.

For $J/\psi\to VP$, since the mass of $J/\psi$ is far below the
$D\bar{D}$ threshold, the open charm effects via the intermediate
meson loops are negligibly small. The overall branching ratios for
$J/\psi\to VP$ can be well described by the SOZI and EM transitions
with a proper phase~\cite{Li:2007ky,Zhao:2006gw}.

We emphasize that the EM transitions also play an important role.
This can be easily recognized by the observation that the
isospin-violating decay channels, $J/\psi$ and $\psi(3686)\to
\rho\eta$ and $\rho\eta^\prime$ have branching ratios compatible
with those isospin conserved ones such as $\omega\eta$,
$\omega\eta^\prime$, $\phi\eta$ and
$\phi\eta^\prime$~\cite{Amsler:2008zzb}. These channels are found
respect ``12\% rule" pretty well and can be well understood in
vector meson dominance (VMD) model~\cite{Li:2007ky,Zhao:2006gw}.

%%%%%%%%%%%%
\section{Further evidences for intermediate meson loops}
%\label{}
\nin
%%%%%%%%%%%%

As a natural mechanism for evading the OZI rule and HSR, and a
rather general non-perturbative scenario in the charmonium energy
region, the intermediate meson loops can also be examined in
different exclusive decay processes, e.g. $\chi_{c1}\to VV$ and
$\chi_{c2}\to VP$. A detailed study of these two decays can be found
in Ref.~\cite{Liu:2009vv}. We summarize the results as follows:

i) Since $\chi_{c0,1,2}$ are $P$-wave states, the short-distance
transition probes the first derivative of the $c\bar{c}$
wavefunction at the origin, which however, would be suppressed in
$\chi_{c1}\to VV$ and $\chi_{c2}\to VP$ due to the HSR.

ii) Experimentally, the branching ratio of $\chi_{c1}\to
K^{*0}\bar{K}^{*0}$ is at order of $10^{-3}$~\cite{Amsler:2008zzb},
which appears to be sizeable, and implies a violation of the HSR.

iii) The introduction of the intermediate meson loops provides a
dynamic mechanism in this process. By annihilating the $c\bar{c}$ at
long distance via the intermediate meson loops, the helicity
selection rule can be evaded.

iv) For $\chi_{c1}\to VV$, since the intermediate $D_s$ and $D_s^*$
pair has higher mass threshold, the production of the $\phi\phi$
will be relatively suppressed in comparison with the non-strange
$\rho\pi$ and $\omega\omega$ apart from the final-state phase space
differences. This corresponds to the flavor symmetry breaking at
leading order.

v) For $\chi_{c2}\to VP$, significant $U$-spin symmetry breaking
implies a larger branching ratio for $\chi_{c2}\to K^*\bar{K}+c.c.$
than for $\chi_{c2}\to \rho\pi$. This prediction can be examined by
BESIII experiment.

In brief, we find that the intermediate hadron loop transitions
provide a natural mechanism for the evasion of the helicity
selection rule as a long-distance transition. It could also be a
mechanism for several existing puzzles in the charmonium energy
region. Systematic studies of various processes may allow us to put
together pieces of information and gain much deeper insights into
the underlying dynamics.

%%%%%%%%%%%%%%%%%%%%%%%%%%%
\section*{Acknowledgements}
\nin
%%%%%%%%%%%%%%%%
The author thanks G. Li, X.-H. Liu, and Y.-J. Zhang for
collaborations on the reported topics. This work is supported, in
part, by the National Natural Science Foundation of China (Grants
No. 10675131 and 10491306), Chinese Academy of Sciences
(KJCX3-SYW-N2), and Ministry of Science and Technology of China
(2009CB825200).



%% References


%\bibliographystyle{elsarticle-num}


%%%%%%%%%%%%%%%%%%%%
%\vfill\eject

%\input{bib_sample}

\end{document}